# Objectivity Data Migration

M. Nowak, K. Nienartowicz, A. Valassi, M. Lübeck, D. Geppert
*CERN, CH-1211 Geneva 23, Switzerland*

In this article we describe the migration of event data collected by the COMPASS and HARP experiments at CERN. Together these experiments have over 300TB of physics data stored in Objectivity/DB that had to be transferred to a new data management system by the end of Q1 2003 and Q2 2003 respectively. To achieve this, data needed to be processed with a rate close to 100MB/s, employing 14 tape drives and a cluster of 30 Linux servers. The new persistency solution to accommodate the data is built upon relational databases for metadata storage and standard "flat" files for the event data. The databases contain collections of $10^9$ events and allow generic queries or direct navigational access to the data, preserving the original C++ user API. The central data repository at CERN is implemented using several Oracle9i servers on Linux and the CERN Mass Storage System CASTOR.

## 1. INTRODUCTION

The history of Objectivity/DB at CERN started in 1995, when it was introduced by the RD45 project as a candidate system for managing the data of the LHC experiments. Objectivity is a fully object-oriented database management system (ODBMS) that offers strong bindings to C++ and Java programming languages and scalability to the Petabyte ($10^{15}$ bytes) range. These features make it well suited to handle physics events data. Some of the LHC experiments built their early software frameworks using Objectivity as the object persistency mechanism and used it in various data challenges and simulated events processing. The interest shown attracted also several pre-LHC experiments which began using Objectivity in production. In some cases the volume of physics data stored in Objectivity was very large.

In mid-nineties the object database market was growing quickly and predictions were made that at the time of the LHC startup ODBMS systems would be commodity software widely supported by industry. However, after 2000 it became apparent that the pure ODBMS market was not developing as predicted, while the traditional relational database products started to incorporate features that allowed the building of very large databases (VLDB) from applications written in C++ and Java in a way similar to ODBMS. Around 2001 the LHC experiments began changing their persistency baseline in favour of alternative solutions, and eventually decided to abandon Objectivity/DB. As consequence, the maintenance contract between CERN and Objectivity was not prolonged beyond 2001.

The end of the maintenance contract did not mean an immediate stop of Objectivity/DB usage. Based on the existing perpetual licenses, all existing users would still be able to run their software on the supported compiler/platform combination. At CERN, the latest versions included:

- Objectivity 6.1.3, g++2.95.2, RedHat 6.x
- Objectivity 6.1.3, CC 5, Solaris 7/8

However, there would be no support from the company in the form of patches, upgrades and bug fixes. Furthermore, the retirement of RedHat 6.x platform scheduled for middle 2003 drew a final line after which CERN could no longer effectively support Objectivity/DB applications.

The date of phasing out Objectivity support at CERN was agreed with the experiments and set to July 2003. The data stored in Objectivity Federations which would be needed after the end of Objectivity service had to be migrated to a new storage solution.

This paper describes the migration of Objectivity data of two CERN SPS experiments: COMPASS and HARP, with a combined data volume above 300TB. The migration carried out by the CERN Database group, with help from the involved experiments and other groups in CERN's IT division.

## 2. DESCRIPTION OF THE MIGRATION PROJECT

The steps necessary to perform a successful data migration can be summarized in the following list:

- Identify data sets to be migrated
- Identify a new storage technology
- Design a new storage system
- Develop the migration software and hardware setup
- Migrate the data
- Adapt experiments' software to the new storage system
- Validate migration results

The steps can be grouped into three stages of the project:

- Preparation
- Migration
- Validation and adaptation

In our case the preparation phase took a form of R&D activity and required the most effort. While planning for migration started in summer 2002 as a part time job, the software development efforts soon become nearly a full time occupation for the rest of the year for 3 people. An important factor was also the fact that a lot of work had been done in advance during the investigation of the suitability of object-relational databases for storing physics data [3].

The migration itself could be compared to a production activity. It was performed during the winter of 2002/2003 and took about 11 weeks. 2 persons supervised the migration, that required a lot of attention in the early phase, but became almost automatic in the second half.

The last stage of software adaptation and validation of the migration results was done in spring 2003. It required several weeks of work from one person from the database group, and another person from the involved experiment.

## 3.  PREPARING FOR THE MIGRATION

### 3.1.  Objectivity Federations at CERN

In 2002, both the LHC and pre-LHC experiments at CERN were using Objectivity-based persistent storage. When considering their data for migration to a different storage technology, the most important factor was if access to this data would be still required after Q2 2003, after the official Objectivity support ended.

In case of the LHC experiments, which had no real physics events yet, but only test and simulated ones, there was no need to preserve the data for an extended time period. The existing databases would stay operational as long as possible, and afterwards they would be simply deleted – made redundant by more recent simulations using upgraded software. Other LHC databases, containing important "non-physics" data (e.g. measurement data from the detector construction process), would be migrated to a purely relational storage by the experiments themselves.

The situation was very different for the pre-LHC experiments, which had real production data in 2002. Using again the criteria of data access requirement for 2004 or later, two experiments were identified: COMPASS and HARP. Both of them were taking data in 2001 and 2002, and COMPASS would continue through 2003 possibly until 2005.

Of the two experiments, COMPASS was collecting a much larger raw data volume – about 300TB per year, with exception of 2001 when it only had ~25TB. Near the end of run the total COMPASS raw and reconstructed data size is expected to surpass 1PB.

At the start of the migration (end 2002) COMPASS had 300TB of raw data in 12 Objectivity federations.

The other experiment selected for migration, HARP, has finished data taking in October 2002. By that time it has collected ~30TB of data into 2 Objectivity federations. HARP has declared the need to access its data for the next few years.

Given the significant difference in data volume, the migration project was designed primarily with COMPASS data in mind and with the intention to reuse the same mechanism to migrate HARP's data later.

### 3.2.  Source Data Format

The task of migrating data of two different experiments has been simplified by several similarities in their Objectivity-based storage systems. In both cases the raw events were stored in their original, undecoded online DATE format [2]. They were kept as opaque, binary blocks, which made it possible to move them to another storage system without any modifications.

The events were grouped into hierarchical collections, containing small additional amount of metadata and summary physics data. In the case of COMPASS, the hierarchy consisted of Periods, Runs, Chunks and Event Headers.

The highest collection level, Period, corresponds to about 10 days of data taking. Originally, it was supposed to reflect the actual periods in the accelerator operation, but it subsequently diverged. This collection level was implemented as a single Objectivity Federation. Such a division of single years data was enforced by physical limitations of Objectivity v6 regarding the number of files per Federation, but also allowed for database schema changes and more flexible data handling. One Period contains 1000-2000 Runs.

A Run is a collection of all data "chunks" that belong to a given accelerator run. A chunk of data corresponds to a single Objectivity database file with events, with a varying size - for raw data often about 1GB, but always below 2GB due to a limit imposed by other software components. A Run could consist of about 100 raw data chunks. Each time events were reconstructed, there would be a new chunk created with corresponding reconstructed events.

Chunks have been implemented as Objectivity containers, which store event summary data (an event header) and navigational information to retrieve the full event.

While the raw events and their metadata were expressed in a relatively simple Objectivity schema, the reconstructed events were stored as complex persistent structures.  They were built from several persistent classes and contained embedded arrays of objects. Although it was finally agreed that the reconstructed events would not be migrated, in favour of recreating them with improved algorithms, the new storage system still had to be able to provide persistency for them.

HARP event collections were similar to the COMPASS data structure, with the exception that HARP had no need to divide its Federations into Periods. In place of periods HARP used settings which described a fixed set of detector parameters. Settings contained Runs, which in turn could have one or two data files (so called partial Runs). A Run was further divided into spills.

### 3.3.  Choosing the Persistency Technology

When the LHC experiments decided to change their baseline persistency model, a common project – POOL - was initiated [1]. The goal of the project was to provide persistency for physics data based on a so-called hybrid storage solution: relational databases used as a high level catalogues, and ROOT I/O mechanism for object streaming into files. POOL is supposed to be used in LCG production activities in Q2 2003.

While such timescale were acceptable for the LHC experiments, COMPASS required a new storage system to be operational some time in advance before the 2003 accelerator run. Unless such a system was installed and sufficiently tested before 2003 data taking, COMPASS would have to continue using the existing Objectivity based persistency storage system at least one year longer. To avoid the risk of an additional year of Objectivity support, COMPASS data migration had to be finished before end of March 2003, which in turn meant that the underlying storage system had to be in place at the end of 2002 – almost a year earlier than the first planned POOL production release.

To cope with this aggressive time schedule, the decision was taken to implement a simple, centralized storage system, dedicated to the COMPASS and HARP data models. The system would follow the same hybrid store principle as POOL, but without built-in support for distributed storage or generic object streaming. However, at the request of the experiments, the system would retain one of the basic features of the Objectivity/DB – navigational access to individual events in the store.

The navigational access to events was realized by implementing an event catalog in a relational database. We decided to use Oracle9i – Oracle because its long history and support at CERN; version 9 because of scalability features and C++ API. No Oracle-specific extensions have been used in the database schema to minimize dependency on a particular vendor and facilitate porting to other databases, if required.

The actual events are kept in regular files, in their original binary format (DATE), which is opaque to the database. Access to event data can be realized by first querying the metadata database about a set of event fulfilling the given criteria. The database responds with the navigational information specifying file names and event locations in these files. The application can then transparently read the event data using appropriate libraries, even if the files have been archived to tape.

A special approach was required to provide persistency for the reconstructed events. In Objectivity, they were stored as persistent C++ objects and there was no readily available way to write them to regular files. As their expected combined size was comparable to the raw data size, it would not be possible to store them entirely in a relational database.

To handle the reconstructed events, the experiments had agreed to provide the streaming code to write them to files. In that way they could be treated as binary data blocks, with the internal structure opaque to the database – similar to the raw events.

### 3.4. New Database Schema

The new relational schema for the event metadata has been design to reflect the hierarchical event collection structure. As presented on Figure 1, the hierarchy has 3 levels: Runs, Files (or chunks) and Event Headers.

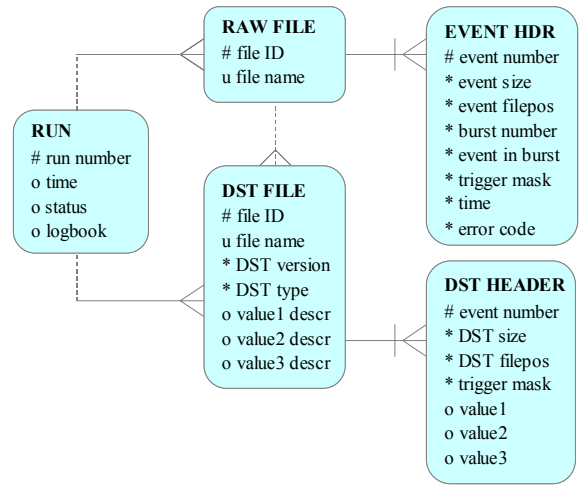

Figure 1: Relational schema for the new storage system

The most important table is the Events table, due to its enormous (in term on relational databases) size. This table is index-organized, as the table rows are rather small and creating a separate index would introduce more storage overhead. Events are uniquely identified by their event number and file ID (assigned internally by the database). Using file ID is preferred to using the Run number, because most data processing is done with file level granularity. The Run/Event number pair can still be used if desired, by joining the Run and File tables, which is done transparently the by experiment's data access libraries.

## 4. DATA MIGRATION

### 4.1. Source Data Sets

At the start of the migration COMPASS and HARP had the following Objectivity data:

#### 4.1.1. COMPASS Data
- 12 Federations
- 300,000 database files stored on 3450 tapes
- 6.1 billion events
- 300TB total data volume

#### 4.1.2. Harp Data
- 2 Federations
- 30,000 database files stored on 367 tapes

- 760 million event
- 30TB data volume

## 4.2.  Migration Process

In general, the process of migrating data from one persistency system to another can be broken down into 3 basic functions: reading, converting and writing out the results.

### 4.2.1. Reading

All Objectivity database files containing raw data were managed by the CERN mass storage system – Castor. In normal operation mode, an Objectivity application would connect to an Objectivity/Castor server (called also an AMS/MSS interface) on a remote disk and request access to a given database file. The server would read the file from tape and store it in a local disk buffer, sending the required data blocks to the application. Every data block had to be requested separately.

This way random access to the data in MSS was transparent to the general user applications, but rather inefficient for sequential reading of the entire data store. The inefficiency comes from two factors:

- Reading individual data blocks through a network with substantial latency generates wait states for packets round-trip. On the CERN LAN, reading through network can be up to 3 times slower than reading locally.
- Reading individual files from tape would result in about 100 mounts to read the entire tape. This would slow the read speed at least 2 times, assuming the requests were not waiting in the tape queue.

The migration system setup has been designed to avoid both above-mentioned problems, i.e. to read entire tape in one go and to access files locally.

### 4.2.2. Format Conversion

The data format conversion software has been written with a maximum reuse of code in mind. The reading part already existed whilst the writing part had to be implemented from scratch, but with the intention that it would later be integrated with the experiments' frameworks and thus allow them to populate the store with new data.

The actual data conversion during the migration was minimal. The raw event data were extracted from Objectivity persistent objects as a binary block, and passed to the writing routines in this format. Event summary data was produced on the fly.

### 4.2.3. Writing

As the result of data conversion two types of output data streams were created: relational metadata and raw events in pure DATE format. The events were written directly to Castor, using the Castor POSIX compliant C API library – RFIO (remote file I/O). There was one output file created for every source database file.

The metadata was written to an Oracle database using Oracle C++ API (part of the OCCI feature). One row with summary data and navigational information was written per event. The inserts to the database were grouped into sets of 1000 to achieve the required performance – 2500 rows per second. During earlier tests, grouping of inserts increased the maximum performance of the database by a factor of almost 50.

The database inserts were committed after each database file (chunk) had been migrated. In case of error, before or during commit, the transaction was rolled back and the output file in Castor deleted. Another attempt to migrate this file could be undertaken any time later.

### 4.2.4. Concurrent Processing

It is important to note that the schema used by both experiments to store their raw data allowed for a high level of concurrency when processing data "chunks" (or files). This feature was introduced to facilitate data processing on large CPU farms, and it had been essential for the migration process as well.

## 4.3.  Migration System Setup

### 4.3.1. Hardware

From the beginning of the migration planning it was known that only the standard CERN hardware would be available for assembling the migration system. The typical CERN Linux disk server specification is as follows:

- Double CPU P3 1GHz system, 1GB RAM
- 5x100GB mirrored IDE disks
- Gigabit Ethernet

To achieve the best performance when reading Objectivity database files, all files stored on a given tape had to be retrieved from Castor together, in a single operation, and stored on a local disk. As the biggest tapes were over 100GB in size (before compression), the local Castor disk buffer had to be built from 2 disks resulting in a 200GB pool.

To allow for the continuous migration of data, a given node had to read new tapes and convert data at the same time. For that purpose a second Castor disk buffer of the same size was configured on each machine. Two processing tasks were running all the time, one reading from tape into one of the two disk pools, the other converting data that has been prepared for it earlier in the other pool. Once both tasks had finished a tape they were processing, they would switch pools. The setup guaranteed that a given disk was never used for reading and writing at the same time. Test performed earlier had shown a visible performance drop when there was only one bigger disk buffer used for both reading and writing operations.

The remaining disk was dedicated for processing logs and for a backup copy of the metadata inserted into to database (in fact, there was never a need to use this extra copy).

The migrated data was written into a remote Oracle database and into Castor files. There was a single large output disk buffer provided for the output files. The output pool was entirely under Castor control and the mass storage system took care of copying files from this pool to tape and managed free space in it.

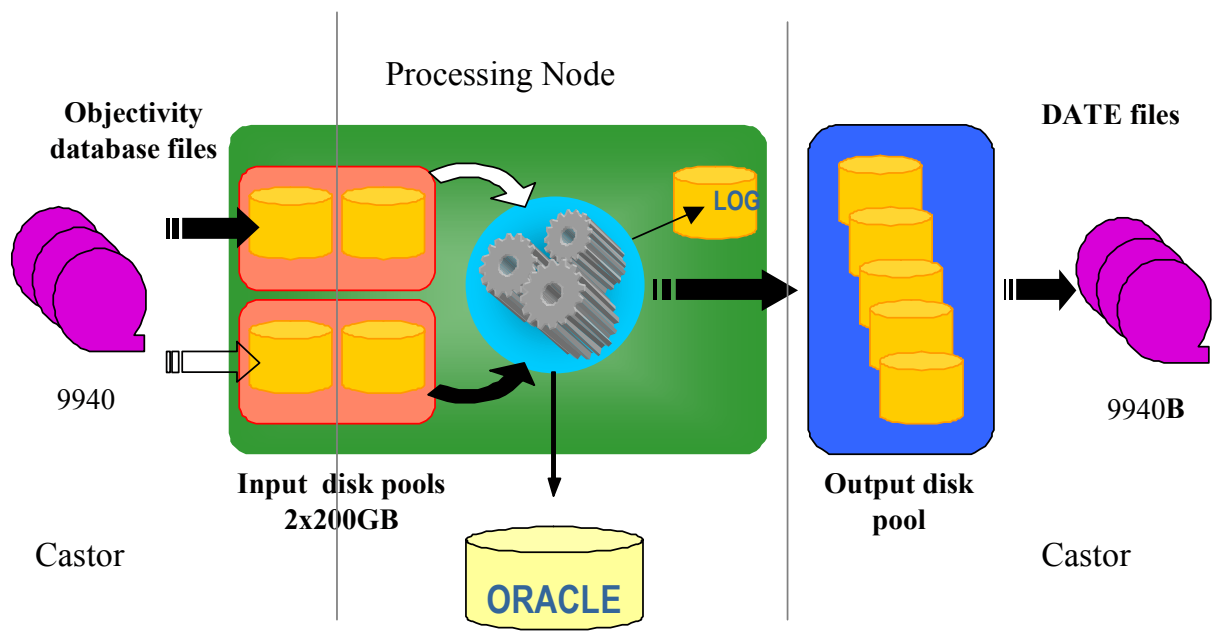

Figure 2:Fisheye view on the migration system, with the focus on a single worker node. Left and right sides of the picture correspond to the tapes and disk buffers under the control of Castor. The central part represents the area of conversion software controlled by the migration manager.

A single migration node configured in the described way could achieve 10MB/s average data throughput, assuming no waiting for input tape drives and no congestions in the output pool. To be able to migrate 300TB in 50 days with this speed, assuming 100% efficiency, 10 processing nodes (with an appropriate output disk pool) would be required. The number of 50 days was used in calculations since the time window for the migration was not much longer than 100 days, and the system efficiency in real life is always lower that 100%.

To sustain data rate of 100MB/s, a number of dedicated input and output tape drives were necessary. As CERN was in the process of installing a new tape drive type – 9940B – it had been decided that the migrated data would be written using the new drive model, to avoid additional media migration soon afterwards. The setup had thus to include a group of 9940A drives for reading and another group of 9940B drives for writing.

The average speed of reading from 9940A tapes and writing to 9940B had been measured for COMPASS data to be 12MB/s and 17MB/s respectively. Therefore 8 to 9 input drives and 6 output drives were needed.

### 4.3.2.Software

The farm of the migration nodes was put under control of the migration manager – software developed for this particular purpose and based on MPI libraries. The manager had its own dedicated database that contained information about all files to be migrated. Using this information the manager was able to distribute workload among the processing nodes. The distribution had been calculated dynamically, taking into account different factors like tape sizes or tape locations vs. available number of tape drives per location. The manager was also able to restart work after an interruption, remembering files that have already been read from tape but not yet migrated.

The manager database stored all information about the progress of the migration. Figures in chapter 4.4 were produced using the web interface to this database. The web interface was also used as the primary tool for overseeing the migration process.

### 4.3.3.List of Resources

The complete list of hardware resources available for the migration changed quite often. The full setup foreseen for the COMPASS was available only for a short time at the end of the migration. During that time, the setup consisted of the following elements:

- 11 processing nodes
- 4.5TB Castor output disk pool
- 3 COMPASS metadata databases
- Migration manager database
- Migration manager
- Castor server (stager)
- 8 dedicated input tape drives (9940A type)
- 10 available output tape drive (9940B type)

The migration manager and Castor server were running on "diskless" nodes. The number of available output tape drives was high, because the drives were freshly installed and the migration was the only user at that time.

HARP migration which followed after the COMPASS one had been finished, had at its disposal a reduced subset of this configuration:

- 4 processing nodes
- 2TB Castor output disk pool
- 1 HARP metadata databases
- Migration manager database
- Migration manager
- Castor server (stager)

## 4.4. Migration Performance

The migration started in December 2003. At the beginning, resources were limited – there were only 4 processing nodes and 1 output tape drive. Most of the hardware became available in the second half of December and the migration speed reached 6TB per day for a short time. At that throughput, the configuration of the mass storage system proved to be inadequate, and the performance dropped almost to zero. This can be seen as the first big dip on Figure 3. After reconfiguring the MSS disk buffers, the migration was restarted and left unattended for the CERN winter break, during which the laboratory was closed. It continued to work until the 1st of January, when a bug in the handling of dates stopped it for several days. After the reopening of the laboratory and fixing the date bug, the migration continued with very uneven performance. Throughout January, many unexpected problems were encountered, including tape shortage, MSS problems, MSS unavailability during relocation, power cuts and some deficiencies in the migration software itself. The overall efficiency was about 2/3 of the planned speed.

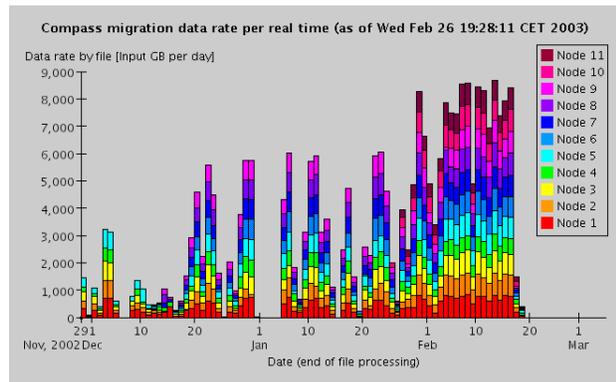

Figure 3: COMPASS migration performance

By February the hardware setup had been increased to 11 processing nodes and the migration software had become much more robust. This increased the throughput to over 8TB per day, and allowed almost uninterrupted processing.

The COMPASS raw data migration was finished on the 19th of February and took in total 10 weeks to complete.

The HARP migration started soon afterwards, but with reduced resources. Only 4 processing nodes were used to migrate the HARP data, but as the volume was only 10% of that of COMPASS, the migration took only 2 weeks. The hardware setup and the core of the migration software was exactly the same as before and by that time well debugged and understood, so no interruptions were encountered. The peak migration speed reached 2.5TB per day, as show on Figure 4.

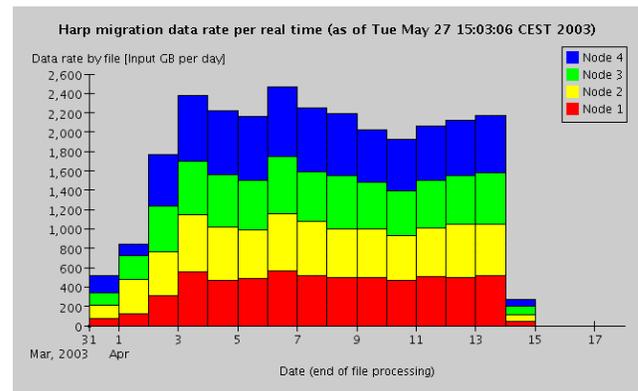

Figure 4: HARP migration performance

## 5. THE NEW STORAGE SYSTEMS

Since the completion of the data migration, COMPASS has moved all data processing to the new, hybrid storage system. Performance comparisons between the old and the new system (see Figure 5) show that the new one can deliver data faster, and the difference grows with the number of concurrent users.

At the moment the setup for hosting 2001 and 2002 COMPASS data consists of 3 Oracle database servers, raw data on tapes in Castor and a set of disk servers used as a staging area for Castor files. User applications are executed on the CERN central shared CPU farm.

The size of the event metadata in Oracle right after the migration was in total 6.1 billion rows, taking 335GB of disk space on all three servers. Each server hosts one database instance, which contains 4 Periods, and each Period has a dedicated database account. In that way the raw metadata is divided between 12 database accounts, which decreases the actual size of a single table to about 500 million rows.

The new system is already being used in production and COMPASS has reconstructed a significant part of the raw data. The production is being performed by 400 concurrent processes.

After reconstruction the results are also inserted into the database in a similar way to the raw data. Assuming each event may be reconstructed up to 3 times using different algorithms, the final combined size of the databases can surpass 20 billion rows and 1TB of disk space.

In addition to the raw data account, every Period has another account that owns the metadata of the reconstructed events. A third account that currently does not own any data is used to provide read access to all the data.

The existence of three different accounts per Period serves as the access control and data protection mechanism. The reconstruction applications cannot modify the raw data, and normal users can only read raw and reconstruction events.

The separation of raw and reconstructed events can also simplify data distribution off-site.

At the moment another set of 3 new database servers is being prepared for this year's data taking. The planned number of physics events to be collected in 2003 is slightly larger than in 2002.

The system setup for HARP follows the COMPASS one, but requires only one database server for the metadata. The database takes about 200GB of disk space, as the amount of summary data per event is larger than for COMPASS.

HARP is preparing to start using the new storage system, but at the time of writing the article it has not moved over yet.

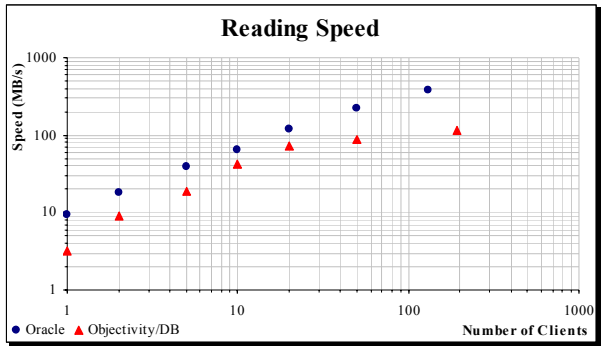

Figure 5: Performance comparison between Objectivity and Oracle based COMPASS data stores. Picture taken from [4]

## 6. SUMMARY

The Objectivity raw data migration was finished successfully, on schedule and with a minimal hardware resource investment. The new hybrid system based on a relational database fulfills the persistency requirement of the experiments providing navigational access to the events, good performance and scalability.

COMPASS, the larger of the two migrated experiments is already using the new system in production and is preparing for 2003 data taking. The original tapes with Objectivity data are being gradually released for reuse.

The migration exercise provides proof of viability of Oracle databases for handling large physics data volumes on Linux systems and commodity PC hardware.

### Acknowledgments

The Database group wishes to thank everybody who contributed to the success of the migration – in particular the Castor team, the Data Services group and the Architecture and Data Challenges group at CERN.